\documentclass[vecphys]{svmult}

\usepackage{makeidx}         
\usepackage{graphicx}        
\usepackage{multicol}        
\usepackage[bottom]{footmisc}
\newcommand{\etal}{{\it et al. }}
\newcommand{\ie}{{\frenchspacing \it i.e. }}

\newcommand\plotone[1]{%
\typeout{Plotone included the file #1}
\centering
\leavevmode
\includegraphics[width={\eps@scaling\columnwidth}]{#1}%
}%
\newcommand\plottwo[2]{{%
\typeout{Plottwo included the files #1 #2}
\centering
\leavevmode
\columnwidth=.45\columnwidth
\includegraphics[width={\eps@scaling\columnwidth}]{#1}%
\hfil
\includegraphics[width={\eps@scaling\columnwidth}]{#2}%
}}
\makeindex             

\begin{document}

\title*{Optical Detection of Galaxy Clusters}
\author{Roy R. Gal\inst{1}}

\institute{University of Virginia, Dept. of Astronomy, PO Box 3818, Charlottesville, VA 22903, USA \\
\texttt{rg8j@virginia.edu}
}

\maketitle

\section{Introduction}
\label{sec1}

Taken literally, galaxy clusters must be comprised of an overdensity
of galaxies. Almost as soon as the debate was settled on whether or not the
``nebulae'' were extragalactic systems, it became clear
that their distribution was not random, with regions of very high
over- and under-density. Thus, from a historical perspective, it is
important to discuss the detection of galaxy clusters through their
galactic components. Today we recognize that galaxies constitute a very
small fraction of the total mass of a cluster, but they are
nevertheless some of the clearest signposts for detection of these
massive systems. Furthermore, the extensive evidence for differential
evolution between galaxies in clusters and the field (discussed at
length elsewhere in these proceedings) means that it is imperative to
quantify the galactic content of clusters.

Perhaps even more importantly, optical detection of galaxy clusters is
now inexpensive both financially and observationally. Large arrays of
CCD detectors on moderate telescopes can be utilized to perform
all-sky surveys with which we can detect clusters to $z\sim0.5$. Using
some of the efficient techniques discussed later in this section, we
can now survey hundreds of square degrees for rich clusters at
redshifts of order unity with 4-meter class telescopes, and similar
surveys, over smaller areas but with larger telescopes are finding
group-mass systems to similar distances.

Looking to the future, ever larger and deeper surveys will permit the
characterization of the cluster population to lower masses and higher
redshifts. Projects such as the Large Synoptic Survey Telescope (LSST)
will map thousands of square degrees to very faint limits (29th
magnitude per square arcsecond) in at least five filters, allowing
the detection of clusters through their weak lensing signal (\ie mass) as well as
the visible galaxies. Ever more efficient cluster-finding algorithms
are also being developed, in an effort to produce catalogs with low
contamination by line-of-sight projections, high completeness, and
well-understood selection functions.

This chapter provides an overview of past and present techniques for
optical detection of galaxy clusters. It follows the progression of
cluster detection techniques through time, allowing readers to
understand the development of the field while explaining the variety
of data and methodologies applied. Within each section we describe the
datasets and algorithms used, pointing out their strengths and
important limitations, especially with respect to the
characterizability of the resulting catalogs. The next section
provides a historical overview of pre-digital, photographic surveys
that formed the basis for most cluster studies until the start of the
twenty-first century. Section three describes the hybrid photo-digital
surveys that created the largest current cluster catalogs. The fourth
section is devoted to fully digital surveys, most
specifically the Sloan Digital Sky Survey and the variety of methods
used for cluster detection. We also describe smaller
surveys, mostly for higher redshift systems. The fifth section gives
an overview of the different algorithms used by these surveys, with an
eye towards future improvements. The concluding section discusses
various tests that remain to be done to fully understand any of the
catalogs produced by these surveys, so that they can be compared to
simulations.

\section{Photographic Cluster Catalogs}
\label{sec2}

Even before astronomers had a full grasp of the distances to other
galaxies, the creators of the earliest catalogs of nebulae recognized
that they were sometimes in spectacular groups. Messier and the
Herschels observed the companions of Andromeda and what we today know
as the Pisces-Perseus supercluster. With the invention of the
wide-field Schmidt telescope, astronomers undertook imaging surveys
covering significant portions of the sky. These quickly revealed some
of the most famous clusters, including Virgo, Coma, and Hydra. The
earliest surveys relied on visual inspection of vast numbers of
photographic plates, usually by a single astronomer. As early as 1938,
Zwicky discussed such a survey based on plates from the $18''$ Schmidt
telescope at Palomar. In 1942, Zwicky and Katz \& Mulders published a
pair of papers presenting the first algorithmic analyses of galaxy
clustering from the Shapley-Ames catalog, using galaxies brighter than
$12.7^m$. Examining counts in cells, cluster morphologies, and
clustering by galaxy type, these surveys laid the foundation for
decades of galaxy cluster studies, but were severely limited by the
very bright magnitude limit of the source material. Nevertheless, many
fundamental properties of galaxy clusters were discovered.  Zwicky,
with his typical prescience, noted that elliptical galaxies are much
more strongly clustered than late-type galaxies (Figure 1), and
attempted to use the structure and velocity dispersions of clusters to
constrain the age of the universe as well as galaxy masses.

\begin{figure}[!ht]
\centering
\includegraphics[height=4cm]{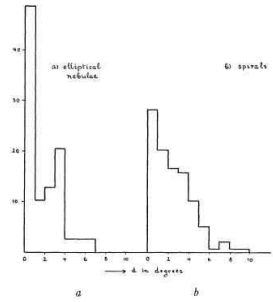}
\caption{The radial distribution of elliptical and spiral ``nebulae'' in the Virgo cluster. The enhanced clustering of elliptical galaxies is apparent, and is used to construct many modern cluster catalogs.}
\end{figure}

However, the true pioneering work in this field did not come until
1957, upon the publication of a catalog of galaxy clusters produced by
George Abell as his Caltech Ph.D. thesis, which appeared in the
literature the following year (Abell 1958). Zwicky followed suit a
decade later, with his voluminous Catalogue of Galaxies and of
Clusters of Galaxies (Zwicky, Herzog \& Wild 1968). However, Abell's
catalog remained the most cited and utilized resource for both galaxy
population and cosmological studies with clusters for over forty
years.  Abell used the red plates of the first National
Geographic-Palomar Observatory Sky Survey. These plates, each spanning
$\sim6^{\circ}$ on a side, covered the entire Northern sky, to a
magnitude limit of $m_r\sim20$. His extraordinary work required the
visual measurement and cataloging of hundreds of thousands of galaxies
To select clusters, Abell applied a number of criteria in an
attempt to produce a fairly homogeneous catalog. He required a minimum
number of galaxies within two magnitudes of the third brightest galaxy
in a cluster ($m_3+2$), a fixed physical size within which galaxies
were to be counted, a maximum and minimum distance to the clusters,
and a minimum galactic latitude to avoid obscuration by interstellar
dust. The resulting catalog, consisting of 1,682 clusters in the
statistical sample, remained the only such resource until 1989. In
that year, Abell, Corwin \& Olowin (hereafter ACO) published an
improved and expanded catalog, now including the Southern sky. These
catalogs have been the foundation for many cosmological studies over
the last four decades, even with serious questions about their
reliability. Despite the numerical criteria laid out to define
clusters in the Abell and ACO catalogs, their reliance on the human
eye and use of older technology and a single filter led to various
biases. These include a bias towards centrally-concentrated clusters
(especially those with cD galaxies), a relatively low redshift cutoff
($z \sim0.15$; Bahcall \& Soneira 1983), and strong plate-to-plate
sensitivity variations. Photometric errors and other inhomogeneities
in the Abell catalog (Sutherland 1988, Efstathiou \etal 1992), as well
as projection effects (Lucey 1983, Katgert \etal 1996) are a serious
and difficult-to-quantify issue. These resulted in early
findings of excess large-scale power in the angular correlation
function (Bahcall \& Soneira 1983), and later attempts to disentangle
these issues relied on models to decontaminate the catalog (Sutherland
1988, Olivier \etal 1990). The extent of these effects is also
surprisingly unknown; measures of completeness and contamination in
the Abell catalog disagree by factors of a few. For instance, Miller,
Batuski \& Slinglend (1999) claim that under- or over-estimation of
richness is not a significant problem, whereas van Haarlem, Frenk \&
White (1997) suggest that one-third of Abell clusters have incorrect
richnesses, and that one-third of rich ($R\ge1$) clusters are
missed. Unfortunately, some of these problems will plague any
optically selected cluster sample, but objective
selection criteria and a strong statistical understanding of the
catalog can mitigate their effects.

In addition to the Zwicky and Abell catalogs, a few others based on
plate material have also been produced, such as Shectman (1985), from
the galaxy counts of Shane \& Wirtanen (1954), and a search for more
distant clusters carried out on plates from the Palomar $200''$ by
Gunn, Hoessel \& Oke (1986; hereafter GHO). None of these achieved the
level of popularity of the Abell catalog, although the GHO survey was
one of the first to detect a significant number of clusters at
moderate to high redshifts ($0.15<z<0.9$), and remains in use to this
day.

\section{Hybrid Photo-Digital Surveys}
\label{sec3}

Only in the past ten years has it become possible to utilize the
objectivity of computational algorithms in the search for galaxy
clusters. These more modern studies required that plates be digitized,
so that the data are in machine readable form. Alternatively, the data
had to be digital in origin, coming from CCD cameras. Unfortunately,
this latter option provided only small area coverage, so the hybrid
technology of digitized plate surveys blossomed into a cottage
industry, with numerous catalogs being produced in the past
decade. All such catalogs relied on two fundamental data sets: the
Southern Sky Survey plates, scanned with the Automatic Plate Measuring
(APM) machine (Maddox \etal 1990) or COSMOS scanner (to produce the
Edinburgh/Durham Southern Galaxy Catalog / EDSGC, Heydon-Dumbleton,
Collins \& MacGillivray 1989), and the POSS-I, scanned by the APS
group (Pennington \etal 1993).  The first objective catalog produced
was the Edinburgh/Durham Cluster Catalog (EDCC, Lumsden \etal 1992),
which covered 0.5 sr ($\sim 1,600$ square degrees) around the South
Galactic Pole (SGP). Later, the APM cluster catalog was created by
applying Abell-like criteria to select overdensities from the galaxy
catalogs, and is discussed in detail in Dalton \etal (1997).  More
recent surveys, such as the EDCCII (Bramel, Nichol \& Pope 2000) did
not achieve the large area coverage of DPOSS (see below), and perhaps
more importantly, are not nearly as deep. For instance, the EDCCII's
limiting magnitude is $b_J=20.5$. For an $L_*$ elliptical this
corresponds to a limiting redshift of $z\sim0.23$. The work by Odewahn
\& Aldering (1995), based on the POSS-I, provided a Northern sky
example of such a catalog, while utilizing additional information
(namely galaxy morphology). Some initial work on this problem, using
higher quality POSS-II data, was performed by Picard (1991) in his
thesis.

The largest, most recent, and likely the last photo-digital cluster
survey is the Northern Sky Optical Survey (NoSOCS; Gal \etal 2000, 2003,
2006; Lopes \etal 2004). This survey relies on galaxy catalogs
created from scans of the second generation Palomar Sky Survey
plates. The POSS-II (Reid \etal 1991) covers the entire northern sky
($\delta > -3^\circ$) with 897 overlapping fields (each $6.5^\circ$
square, with $5^\circ$ spacings), and, unlike the old POSS-I, has no
gaps in the coverage.  Approximately half of the survey area is
covered at least twice in each band, due to plate overlaps.  Plates
are taken in three bands: blue-green, IIIa-$J$ + GG395, $\lambda_{\rm
eff} \sim480nm$; red, IIIa-$F$ + RG610, $\lambda_{\rm eff} \sim650nm$;
and very near-IR, IV-$N$ + RG9, $\lambda_{\rm eff} \sim850nm$.
Typical limiting magnitudes reached are $B_J \sim22.5$, $R_F
\sim20.8$, and $I_N \sim19.5$, \ie, $\sim1^m - 1.5^m$ deeper than
POSS-I.  The image quality is improved relative to POSS-I, and is
comparable to the southern photographic sky surveys. The original
survey plates are digitized at STScI, using modified PDS scanners
(Lasker 1996). The plates are scanned with $15\mu$ ($1.0''$) pixels,
in rasters of 23,040 square, giving $\sim1$ GB/plate, or $\sim3$ TB of
pixel data total for the entire digital survey. The digital scans
are processed, calibrated, and cataloged, with detection of all
objects down to the survey limit, and star/galaxy classifications
accurate to 90\% or better down to $\sim1^m$ above the detection limit
(Odewahn \etal 2004). They are photometrically calibrated using
extensive CCD observations of Abell clusters (Gal \etal 2004a). 

The resulting galaxy catalogs are used as an input to an adaptive
kernel galaxy density mapping routine (discussed in \S5), and
photometric redshifts based on galaxy colors are calculated, along
with cluster richnesses in a fixed absolute luminosity interval.  The
NoSOCS survey utilizes $F$ (red) plates, with a limiting magnitude of
$m_r=20$. This corresponds to a limiting redshift of 0.33 for an $L_*$
elliptical galaxy. Because of the increase in $g-r$ color with
redshift, the APM would have to go as deep as $b_J=22.0$ to reach the
same redshift from their data for early type galaxies. Similarly, even
at lower redshift, this implies that DPOSS can see $\sim0.5^m-1^m$
deeper in the cluster luminosity functions. Additionally, NoSOCS uses
at least one color (two filters), and a significantly increased amount
of CCD photometric calibration data. The final catalog covers 11,733
square degrees, with nearly 16,000 candidate clusters (Figure 2),
extending to $z\sim0.3$, making it the largest such resource in
existence. However, new CCD surveys, discussed in the next section,
are about to surpass even this benchmark.

\begin{figure}[!ht]
\centering
\includegraphics[height=6cm]{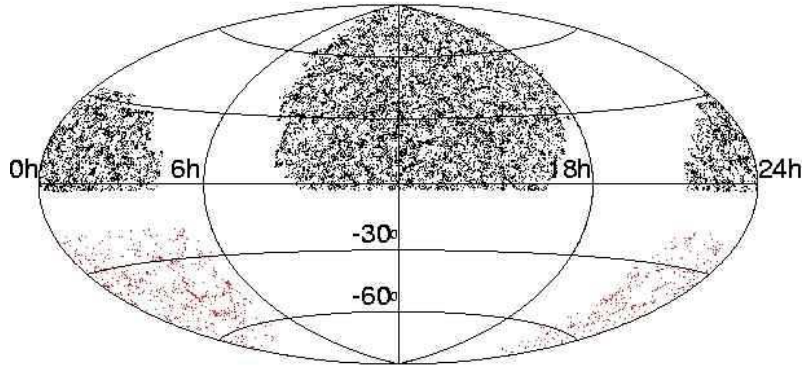}
\caption{The sky distribution of NoSOCS (northern sky) and APM (southern sky) candidate clusters in equatorial coordinates. The much higher density of NoSOCS is due to its deeper photometry and lower richness limit.}
\end{figure}

\section{Digital CCD Surveys}
\label{sec4}

With the advent of charge-coupled devices (CCDs), fully digital
imaging in astronomy became a reality. These detectors provided an
order-of-magnitude increase in sensitivity, linear response to light,
small pixel size, stability, and much easier calibration. The main
drawback relative to photographic plates was (and remains) their small
physical size, which permits only a small area (of order $10'$) to be
imaged by a typical $2048^2$ pixel detector. As detector sizes grew,
and it became possible to build multi-detector arrays covering large
areas, it became apparent that new sky surveys with this modern
technology could be created, far surpassing their photographic
precursors. Unfortunately, in the 1990s most modern telescopes did not
provide large enough fields-of-view, and building a sufficiently large
detector array to efficiently map thousands of square degrees was
still challenging.

Nevertheless, realizing the vast scientific potential of such a
survey, an international collaboration embarked on the Sloan Digital
Sky Survey (SDSS, York \etal 2000), which included construction of a
specialized 2.5 meter telescope, a camera with a mosaic of 30 CCDs, a
640-fiber multi-object spectrograph, a novel observing strategy, and
automated pipelines for survey operations and data processing. Main
survey operations were completed in the fall of 2005, with over 8,000
square degrees of the northern sky image in five filters to a depth of
$r'\sim22.2$ with calibration accurate to $\sim2-3\%$, as well as
spectroscopy of nearly one million objects.

With such a rich dataset, many groups both internal and external to
the SDSS collaboration have generated a variety of cluster catalogs,
from both the photometric and the spectroscopic catalogs, using
techniques including:
\begin{enumerate}
\item Voronoi Tessellation (Kim \etal 2002)
\item Overdensities in both spatial and color space (maxBCG, Annis \etal 1999)
\item Subdividing by color and making density maps (Cut-and-Enhance, Goto \etal 2002)
\item The Matched Filter and its variants (Kim \etal 2002)
\item Surface brightness enhancements (Bartelmann \etal 2002, Zaritsky \etal 1997, 2002)
\item Overdensities in position and color spaces, including redshifts (C4; Miller \etal 2005)
\end{enumerate}
These techniques are described in more detail in \S5. Each method
generates a different catalog, and early attempts to compare them have
shown not only that the catalogs are quite distinct, but also that
comparison of two photometrically-derived catalogs, even from the same
galaxy catalogs, is not straightforward (Bahcall \etal 2003).

In addition to the SDSS, smaller areas, to much higher redshift, have
been covered by numerous deep CCD imaging surveys. Notable examples
include the Palomar Distant Cluster Survey (PDCS, Postman \etal 1996),
the ESO Imaging Survey (EIS, Lobo \etal 2000), Zaritsky \etal (1997),
and many others. None of these surveys provide the angular coverage
necessary for large-scale structure and cosmology studies, and are
specifically designed to find rich clusters at high redshift. The
largest such survey to date is the Red Sequence Cluster Survey (RCS,
Gladders \etal 2005), based on moderately deep two-band imaging using
the CFH12K mosaic camera on the CFHT 3.6m telescope, covers $\sim100$
square degrees. This area coverage makes it comparable to or larger
than X-ray surveys designed to detect clusters at $z\sim1$. The use of
the red sequence of early-type galaxies makes this a very efficient
survey, and the methodology is described is \S5.

\section{Algorithms}

From our earlier discussion, it is obvious that many different
mathematical and methodological choices must be made when embarking on
an optical cluster survey. Regardless of the dataset and algorithms
used, a few simple rules should be followed to produce a catalog that
is useful for statistical studies of galaxy populations and for
cosmological tests:
\begin{enumerate}
\item Cluster detection should be performed by an objective, automated
algorithm to minimize human biases and fatigue.
\item The algorithm utilized should impose minimal constraints on the
physical properties of the clusters, to avoid selection biases. If
not, these biases must be properly characterized.
\item The sample selection function must be well-understood, in terms
of both completeness and contamination, as a function of both redshift
and richness. The effects of varying the cluster model on the
determination of these functions must also be known.
\item The catalog should provide basic physical properties for all the
detected clusters, including estimates of their distances and some
mass-related quantity (richness, luminosity, overdensity) such that
specific subsamples can be selected for future study.
\end{enumerate}

This section describes many of the algorithms used to detect clusters
in modern cluster surveys. No single one of these generates an
'optimal' cluster catalog, if such a thing can even be said to
exist. Therefore, I provide some of the strengths and weaknesses of
each technique. In addition to the methods discussed here,
many other variants are possible, and in the future, joint detection
at multiple wavelengths (\ie optical and X-ray, Schuecker \etal 2004)
may yield more complete samples to higher redshifts and lower mass
limits, with less contamination.

\subsection{Counts in Cells}

The earliest cluster catalogs, like those of Abell, utilized a simple
technique of counting galaxies in a fixed magnitude interval, in cells
of a fixed physical or angular size. Indeed, Abell simply used visual recognition of galaxy overdensities, whose properties were then measured {\it ex post facto} in fixed physical cells. This technique was used by Couch
\etal (1991) and Lidman \& Peterson (1996) to detect clusters at
moderate redshifts ($z\sim0.5$), by requiring a specified enhancement, above the mean background, of the galaxy surface density in a given area. This enhancement, called the contrast, is defined as
\begin{equation}
{\sigma_{cl} = {N_{cluster} - N_{field} \over \sigma_{field}}}
\end{equation}
where $N_{cluster}$ is the number of galaxies in the cell
corresponding to the cluster, $N_{field}$ is the mean background
counts and $\sigma_{field}$ is the variance of the field counts for
the same area.  The magnitude range and cell size used are parameters
that must be set based on the photometric survey material and the type
or distance of clusters to be found. For instance, Lidman \& Peterson
(1996) chose these parameters to maximize the contrast above
background for a cluster at $z=0.5$. Using the distribution of cell
counts, one can analytically determine the detection likelihood of a
cluster with a given redshift and richness (assuming a fixed
luminosity function), given a detection threshold.  The false
detection rate is harder, if not impossible, to quantify, without
running the algorithm on a catalog with extensive spectroscopy. This
is true for most of the techniques that rely on photometry alone. It
is also possible to increase the contrast of clusters with the
background by weighting galaxies based on their luminosities and
positions. Galaxies closer to the cluster center are upweighted, while
the luminosity weighting depends on both the cluster and field
luminosity functions, as well as the cluster redshift. This scheme is
similar to that used by the matched filter algorithm, detailed later.

This technique, although straightforward, has numerous
drawbacks. First, it relies on initial visual detection of
overdensities, which are then quantified objectively. Since simple
counts-in-cells methods use the galaxy distribution projected along
the entire line of sight, chance alignments of poorer systems become
more common, increasing the contamination. Optimizing the magnitude
range and cell size for a given redshift reduces the efficiency of
detecting clusters at other redshifts, especially closer ones since
their core radii are much larger. Setting the magnitude range
typically assumes that the cluster galaxy luminosity function at the
redshift of interest is the same as it is today, which is not
true. Furthermore, single band surveys observe different portions of
the rest frame spectrum of galaxies at different redshifts, altering
the relative sensitivity to clusters over the range probed. Finally,
the selection function can only be determined analytically for
circular clusters with fixed luminosity functions. Given these issues,
this technique is inappropriate for modern, deep surveys.

\subsection{Percolation Algorithms}

A majority of current cluster surveys rely on a smoothed map of
projected galaxy density from which peaks are selected (see
below). However, smoothing invariable reduces the amount of
information being used, leading some authors to employ percolation (or
friends-of-friends, FOF) algorithms. In their simplest form, these
techniques link pairs of galaxies that are separated by a distance
less than some threshold (typically related to the mean galaxy
separations). Galaxies that have links in common are then assigned to
the same group; once a group contains more than a specified number of
members, it becomes a candidate cluster. This technique was used by
Dalton \etal (1997) to construct a cluster catalog from APM data.
However, it is not typically used on two dimensional data, because the
results of this method are very sensitive to the linking length, and
can easily combine multiple clusters into long, filamentary
structures. On the other hand, FOF algorithms are very commonly used
for structure finding in three-dimensional data, especially N-body
simulations (Davis \etal 1985, Efstathiou \etal 1988) and redshift
surveys (Huchra \& Geller 1982, Ramella \etal 2002). A variant of this
technique utilizing photometric redshifts has been recently proposed
by Botzler \etal (2004).

\subsection{Simple Smoothing Kernels}

Another objective and automated approach to cluster detection is the
use of a smoothing kernel to generate a continuous density field from
the discrete positions of galaxies in a catalog. For instance,
Shectman (1985) used the galaxy counts of Shane \& Wirtanen in $10'$
bins, smoothed with a very simple weighting kernel. A minimum number
of galaxies within this smoothed region (in this case, 20) were then
required to detect a cluster. This type of kernel is fixed in angular
size and thus does not smooth clusters at different redshifts with
consistent {\em physical} radii, making its sensitivity highly
redshift dependent. Similarly, it uses the full projected galaxy
distribution (much as Abell did), and is thus insensitive to the different parts of the LF sampled at different redshifts.

\subsection{The Adaptive Kernel}
A slightly more sophisticated technique is to use an adaptive smoothing kernel (Silverman 1986). This technique uses a two--stage process to produce a density map. First, at each point $t$, it produces a pilot estimate $f(t)$ of the galaxy density at each point in the map. Based on this pilot estimate, it then applies a smoothing kernel whose size changes as a function of the local density, with a smaller kernel at higher density. This is achieved by defining a {\it local bandwidth factor}:
\begin{equation}
{\lambda_i = [f(t)/g]^{-\alpha},}
\end{equation}
where $g$ is the geometric mean of $f(t)$ and $\alpha$ is a sensitivity parameter that sets the variation of kernel size with density. NoSOCS uses a sensitivity parameter $\alpha = 0.5$, which results in a minimally biased final density estimate, and is simultaneously more sensitive to local density variations than a fixed--width kernel (Silverman 1986). This is then used to construct the adaptive kernel estimate:
\begin{equation}
{\hat f(t) = n^{-1}\sum_{i=1}^nh^{-2}\lambda_i^{-2}K\{h^{-2}\lambda_i^{-2}(t-X_i)\}}
\end{equation}
where $h$ is the bandwidth, which is a parameter that must be set based on the survey properties. 

The adaptive kernel was used to generate the Northern Sky Optical
Cluster Survey (Gal \etal 2000, 2003, 2006). The smoothing size (in
their case, $500''$ radius) is set to prevent over--smoothing the cores
of higher redshift ($z\sim0.3$) clusters, while avoiding fragmentation
of most low redshift ($z\sim0.08$) clusters. Because the input galaxy
catalog is relatively shallow, and the redshift range probed is not
very large, it is possible to do this. For deeper surveys, this is not
practical, and therefore this technique cannot be used in its simplest
form. Figure 3 demonstrates example density maps, showing the effect
of varying the initial smoothing window. In this figure, four
simulated clusters are placed into a simulated background,
representing the expected range of detectability in the NoSOCS
survey. There are two clusters at low $z$ (0.08), and two at high $z$
(0.24), with one poor and one rich cluster at each redshift (100 and
333 total members, $N_{gals}=25$ and 80 respectively).

\begin{figure}[!ht]
\centering
\includegraphics[height=8cm]{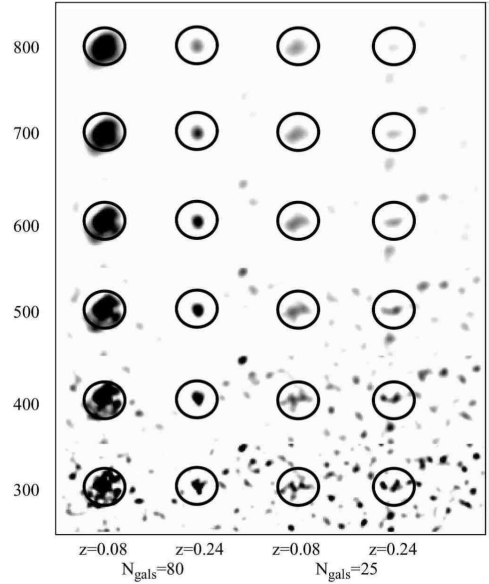}
\caption{The effect of varying the initial smoothing window for the adaptive kernel on cluster appearance. Each panel contains a simulated background with four simulated clusters, as described in the text. The smoothing kernel ranges in size from $300''$ to $800''$ in $100''$ increments. Taken from Gal \etal (2003).}
\end{figure}

After a smooth density map is generated, cluster detection can be
performed analogously to object detection in standard astronomical
images. In NoSOCS, Gal \etal used SExtractor (Bertin \& Arnouts 1996)
to detect density peaks. The tuning of parameters in the detection
step is fundamentally important in such surveys, and can be
accomplished using simulated clusters placed in the observed density
field, from which the completeness and false detection rates can be
determined.  Even so, this method involves many adjustable parameters
(the smoothing kernel size, sensitivity parameter, and all the source
detection parameters) such that it must be optimized with care for the
data being used. Given an end-to-end cluster detection methodology,
one can use simulations to determine the selection function's
dependence on redshift, richness, and other cluster properties (see
Gal \etal 2003 for details). However, the measurement of cluster
richness and redshift are done in a step separate from detection,
using the input galaxy catalogs, further complicating this technique.
The adaptive kernel is very fast and simple to implement, making it
suitable for all--sky surveys, but is only truly useful in situations
where the photometry is poor, and the survey is not very deep, as is
the case for NoSOCS.

\subsection{Surface Brightness Enhancements}
It is not necessary to have photometry for individual galaxies to
detect clusters. A novel but difficult approach is to detect the
localized cumulative surface brightness enhancement due to unresolved
light from galaxies in distant clusters. This method was pioneered by
Zaritsky \etal (1997, 2001), who showed that distant clusters could be
detected using short integration times on small 1-m class
telescopes. However, this method requires extremely accurate
flat-fielding, object subtraction, masking of bright stars. and
excellent data homogeneiety. Once all detected obejcts are removed
from a frame, and nuisance sources such as bright stars masked, the
remaining data is smoothed with a kernel comparable to the size of
clusters at the desired redshift. The completeness and contamination
rates of such a catalog are extremely difficult to model. Thus, this
technique is not necessarily appropriate for generating statistical
catalogs for cosmological tests, but is an excellent, cost-effective
means to find interesting objects for other studies.

\subsection{The Matched Filter}
With accurate photometry, and deeper surveys, one can use more
sophisticated tools for cluster detection. As we will discuss later,
color information is very powerful, but is not always available.
However, even with single--band data, it is possible to simultaneously
use the locations and magnitudes of galaxies. One such method is the
matched filter (Postman \etal 1996), which models the spatial and
luminosity distribution of galaxies in a cluster, and tests how well
galaxies in a given sky region match this model for various
redshifts. As a result, it outputs an estimate of the redshift and
total luminosity of each detected cluster as an integral part of the
detection scheme. Following Postman \etal we can describe, at any location, the distribution of galaxies per unit area and magnitude $D(r,m)$ as a sum of the background and possible cluster contributions:
\begin{equation}
{D(r,m)= b(m) + \Lambda_{cl}P(r/r_c)\phi(m-m^*)}
\end{equation}
Here, D is the number of galaxies per magnitude per arcsec$^2$ at
magnitude $m$ and distance $r$ from a putative cluster center. The
background density is $b(m)$, and the cluster contribution is defined
by a parameter $\Lambda_{cl}$ proportional to its total richness, its
differential luminosity function $\phi(m-m^*)$, and its projected
radial profile $P(r/r_c)$. The parameter $r_c$ is the characteristic
cluster radius, and $m^*$ is the characteristic galaxy luminosity. One
can then construct a likelihood for the data given this model, which
is a function of the parameters $r_c$,$m^*$, and
$\Lambda_{cl}$. Because two of these parameters, especially $m^*$, are
sensitive to the redshift, one obtains an estimated redshift when
maximizing the likelihood relative to this parameter. The algorithm
outputs the richness $\Lambda_{cl}$ at each redshift tested, and thus
provides an integrated estimator of the total cluster richness.  The
luminosity function used by Postman \etal is a Schechter (1976)
function with a power law cutoff applied to the faint end, while they
use a circularly symmetric radial profile with core and cutoff radii
(see their eqn. 19).

Like the adaptive kernel, this method produces density maps on which
source detection must still be run. These maps have a grid size set by
the user, typically of order half the core radius at each redshift
used, with numerous maps for each field, one for each redshift
tested. The goal of the matched filter is to improve the contrast of
clusters above the background, by convolving with an 'optimal' filter,
and also to output redshift and richness estimates. Given a set of
density maps, one can use a variety of detection algorithms to select
peaks. A given cluster is likely to be detected in multiple maps (at
different redshifts) of the same region; its redshift is estimated by
finding the filter redshift at which the peak signal is maximized. By
using multiple photometric bands, one can run this algorithm
separately on each band and improve the reliability of the
catalogs. The richness of a cluster is measured from the density map
corresponding to the cluster redshift, and represents approximately
the equivalent number of $L_*$ galaxies in the cluster.

The matched filter is a very powerful cluster detection technique. It
can handle deep surveys spanning a large redshift range, and provides
redshift and richness measures as an innate part of the procedure. The
selection function can be estimated using simulated clusters, as was
done in significant detail by Postman \etal However, the technique
relies on fixed analytic luminosity functions and radial profiles for
the likelihood estimates. Thus, clusters which have properties
inconsistent with these input functions will be detected at lower
likelihood, if at all. While this is not likely to be an issue at low
to moderate redshifts, as the population of clusters becomes
increasingly merger dominated at $z\sim 0.8$ (Cohn \etal 2005), these
simple representations will fail. Similarly, the cluster and field LF
both evolve with redshift, which can effect the estimated
redshift. Also, as the redshifts and $k$-corrections become large, one
samples a very different region of the LF than at low
redshift. Nevertheless, this remains one of the best cluster detection
techniques for cluster detection in moderately deep surveys.

\subsection{Hybrid and Adaptive Matched Filter}

The matched filter can be extended to include estimated (photometric)
or measured (spectroscopic) redshifts. This extension has been called
the adaptive matched filter (AMF, Kepner \etal 1999). The adaptive
here refers to this method's ability to accept 2-dimensional (positions
and magnitudes), 2.5-dimensional (positions, magnitudes, and estimated
redshifts), and 3-dimensional (positions, magnitudes, and redshifts)
data, adapting to the redshift errors. In implementation, this
technique uses a two--stage method, first maximizing the cluster
likelihood on a coarse grid of locations and redshifts, and then
refining the redshift and richness on a finer grid. Unlike the
standard matched filter, the AMF evaluates the likelihood function at
each galaxy position, and not on a fixed grid for each redshift
interval. Thus, for each galaxy, the output includes a likelihood that
there is a cluster centered on this galaxy, and the estimated
redshift.

The inclusion of photometric redshifts should substantially improve
detection of poor clusters, which is very important since most
galaxies live in poor systems, and these are suspected to be sites for
significant galaxy evolution. However, Kim \etal (2002), using SDSS
data, found that the simple matched filter is more efficient at
detecting faint clusters, while the AMF estimated cluster properties
more accurately. The matched filter performs better for detection
because the significance threshold for finding candidates is redshift
dependent, determined separately for each map produced in different
redshift intervals. The AMF, on the other hand, finds peaks first in
redshift space, and then selects candidates using a universal
threshold. Thus, they propose a hybrid system, using the matched
filter to detect candidate clusters, and the AMF to obtain its
properties.

\subsection{Cut-and-Enhance}

\begin{figure}[!ht]
\centering
\includegraphics[height=6cm]{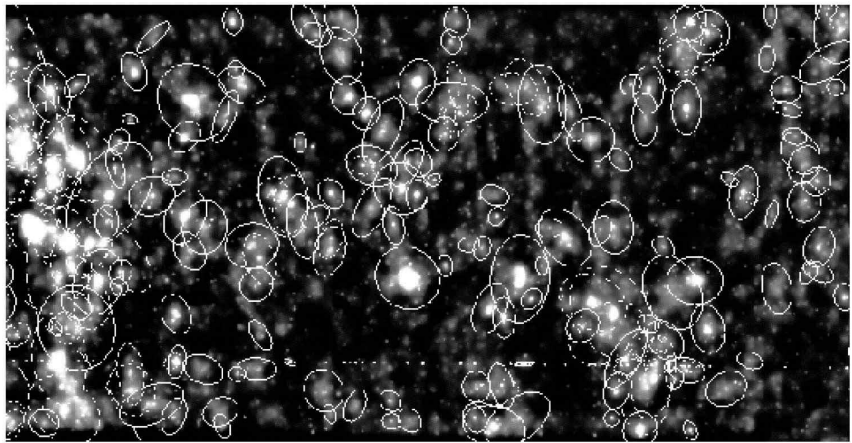}
\caption{An enhanced map of the galaxy distribution in the SDSS Early Data Release, after applying the $g*-r*-i*$ color-color cut. Detected clusters are circled. Taken from Goto \etal (2002).}
\end{figure}

Despite the popularity of matched filter algorithms for cluster
detection, their assumption of a radial profile and luminosity
function are cause for concern. Thus, development of semi-parametric
detection methods remains a vibrant area of research. While the
adaptive kernel described earlier is such a technique, more
sophisticated algorithms are possible, especially with the inclusion
of color information. One such technique is the Cut-and-Enhance method
(Goto \etal 2002), which has been applied to SDSS data. This method
relies on the presence of the red sequence in clusters, applying a
variety of color and color-color cuts to generate galaxy subsamples
which should span different redshift ranges. Within each cut, pairs of
galaxies with separations less than $5'$ are replaced by Gaussian
clouds, which are then summed to generate density maps. In this
technique, the presence of many close pairs (as in a high redshift
cluster) yields a more compact cloud, making it easier to detect, and
thus possibly biasing the catalog against low-$z$ clusters. As with
the AK technique, this method yields a density map on which object
detection must be performed; Goto \etal (2002) utilize
SExtractor. Once potential clusters are detected in the maps made
using the various color cuts, these catalogs must be merged to produce
a single list of candidates. Redshift and richness estimates are
performed {\it a posteriori}, as they are with the AK. Similar to the
AK, there are many tunable parameters which make this method difficult
to optimize.

\subsection{Voronoi Tessellation}

\begin{figure}[!ht]
\centering
\includegraphics[height=9cm]{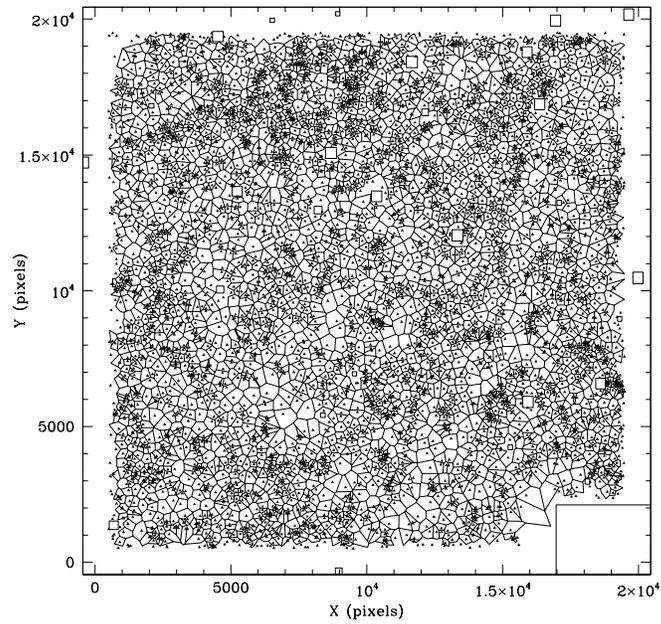}
\caption{Voronoi Tessellation of galaxies with $17.0 \le m_r \le 18.5$ in a DPOSS field. Each triangle represents a
galaxy surrounded by its associated Voronoi cell (indicated by
the polyhedrals).
Excised areas (due to bright objects) are shown as rectangles. Taken from Lopes \etal (2004)}
\end{figure}

Considering a distribution of particles it is possible to define a
characteristic volume associated with each particle.  This is known as
the Voronoi volume, whose radius is of the order of the mean particle
separation.  The complete division of a region into these volumes is
known as Voronoi Tessellation (VT), and it has been applied to a
variety of astronomical problems, and in particular to cluster
detection by Kim \etal (2002) and Ramella \etal (2001). As pointed out
by the latter, one of the main advantages of employing VT to look for
galaxy clusters is that this technique does not distribute the data in
bins, nor does it assume a particular source geometry intrinsic to the
detection process.  The algorithm is thus sensitive to irregular and
elongated structures.

\begin{figure}[!ht]
\centering
\includegraphics[height=8cm]{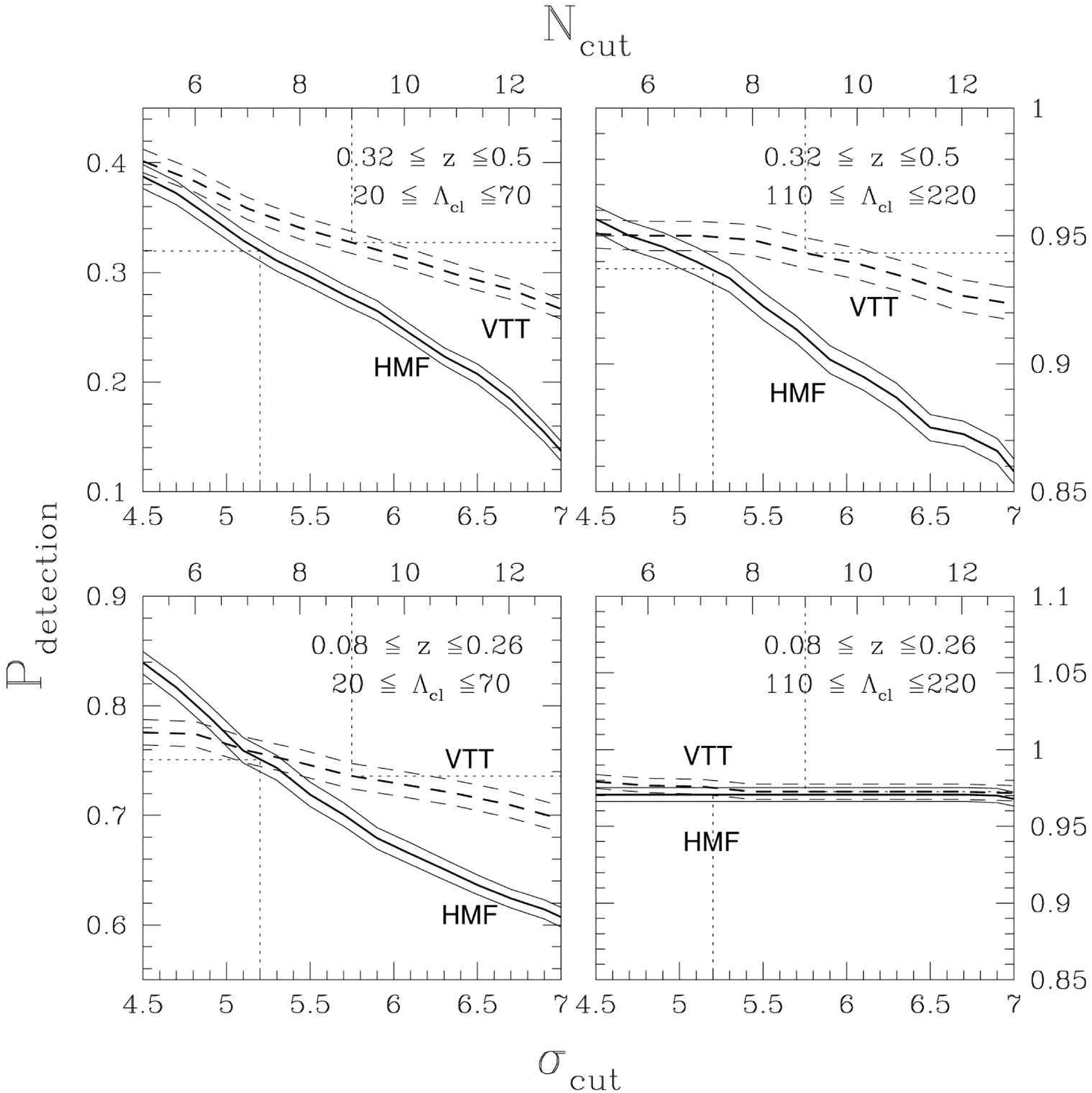}
\caption{The absolute
recovery rates of clusters from the SDSS in four different ranges of cluster
parameters for the HMF (solid line) and the VTT (dashed line). Taken from Kim \etal (2002).}
\end{figure}

The parameter of interest in this case is the galaxy density. When
applying VT to a galaxy catalog, each galaxy is considered as a seed
and has a Voronoi cell associated to it. The area of this cell is
interpreted as the effective area a galaxy occupies in the plane. The
inverse of this area gives the local density at that point. Galaxy
clusters are identified by high density regions, composed of small
adjacent cells, \ie, cells small enough to give a density value higher
than the chosen density threshold.  An example of Voronoi Tessellation
applied to a galaxy catalog for one DPOSS field is
presented in Figure 5. For clarity, we show only galaxies with $17.0
\le m_r \le 18.5$.

\begin{figure}[!ht]
\centering
\includegraphics[height=8cm]{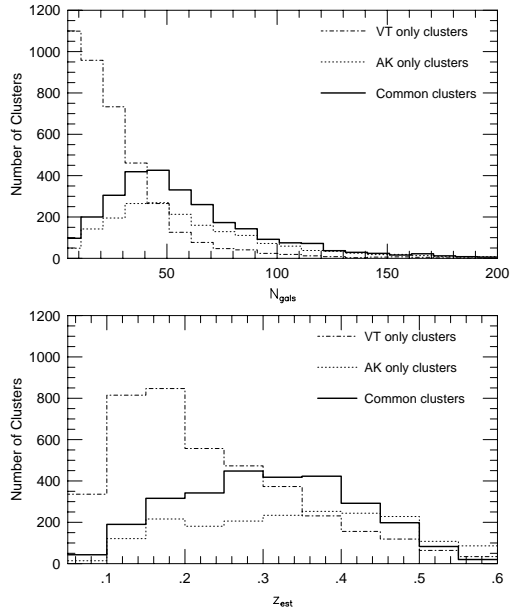}
\caption{ Richness (top) and estimated redshift (bottom) distributions for clusters detected in DPOSS by only the VT code (dash-dotted line), only the AK code (dotted line), and by both methods (heavy solid line). Taken from Lopes \etal (2004).}
\end{figure}

Once such a tessellation is created, candidate clusters are identified
based on two criteria. The first is the density threshold, which is
used to identify fluctuations as significant overdensities over the
background distribution, and is termed the search confidence level
({\bf scl}). The second criterion rejects candidates from the
preliminary list using statistics of Voronoi Tessellation for a
poissonian distribution of particles, by computing the
probability that an overdensity is a random fluctuation.  This is
called the rejection confidence level ({\bf rcl}). Kim \etal (2002)
used the color--magnitude relation for cluster ellipticals to divide
the galaxy catalog into separate redshift bins, and ran the VT code on
each bin. Candidates in each slice are identified by requiring a
minimum number $N_{hdg}$ of galaxies having overdensities $\delta$
greater than some threshold $\delta_c$, within a radius of 0.7$h^{-1}$
Mpc. The candidates originating in different bins are then
cross-correlated to filter out significant overlaps and produce the
final catalog. Ramella \etal (2001) and Lopes \etal (2004) follow a
different approach, as they do not have color information. Instead,
they use the object magnitudes to minimize background/foreground
contamination and enhance the cluster contrast, as follows:

\begin{enumerate}
\item The galaxy catalog is divided into different magnitude bins,
starting at the bright limit of the sample and shifting to progressively
fainter bins. The step size adopted is derived from
the photometric errors of the catalog.

\item The VT code is run using the galaxy catalog for each bin,
resulting in a catalog of cluster candidates associated with each
magnitude slice.

\item The centroid of a cluster candidate detected in different bins
will change due to the statistical noise of the foreground/background
galaxy distribution. Thus, the cluster catalogs from all bins are
cross-matched, and overdensities are merged according
to a set criterion, producing a combined catalog.

\item A minimum number (N$_{min}$) of detections in different
bins is required in order to consider a given fluctuation as a cluster
candidate. N$_{min}$ acts as a final threshold for the whole
procedure.  After this step, the final cluster catalog is complete.
\end{enumerate}

Kim \etal (2002) and Lopes \etal (2004) compare the performance of
their VT algorithms with the HMF and adaptive kernel,
respectively. Figure 6 (taken from Kim \etal 2002) shows the absolute
recovery rates of clusters in four different ranges of cluster
parameters for the HMF (solid line) and the VT (dashed line). Both
algorithms agree very well for clusters with the highest signals
(rich, low redshift), but the VT does slightly better for the
thresholds determined from the uniform background case. Similarly,
Lopes \etal (2004) find that the VT algorithm performs better for
poor, nearby clusters, while the adaptive kernel goes deeper when
detecting rich systems, as seen in Figure 7, where the VT-only
detections are preferentially poor and low redshift, and the AK-only
detections are richer and at high redshift.

\subsection{maxBCG}

The maxBCG algorithm, developed for use on SDSS data (Annis et
al. 2002, Hansen \etal 2005), is another technique that relies on the
small color dispersion of early--type cluster galaxies. The brightest
of the cluster galaxies (BCGs) have predictable colors and magnitudes
out to redshifts of order unity. Unlike many of the other techniques
discussed above, maxBCG does not generate density maps. Instead, it
calculates a likelihood as a function of redshift for {\it each}
galaxy that it is a BCG, based on its colors and the presence of a red
sequence from the surrounding objects. This is calculated as
\begin{equation}
{{\mathcal L}_{max} = max{\mathcal L}(z) ; {\mathcal L}(z)={\mathcal L}_{BCG}+\log N_{gal}}
\end{equation}
where ${\mathcal L}_{BCG}$ is the likelihood, at redshift $z$, that a
galaxy is a BCG, based on its colors and luminosity, and $N_{gal}$ is
the number of galaxies within 1 $h^{-1}$ Mpc with colors and
magnitudes consistent with the red sequence (\ie within 0.1 mag of the
mean BCG color at the redshift being tested). This procedure results
in a maximum likelihood and redshift for each galaxy in the
catalog. The peaks in the ${\mathcal L}_{max}$ distribution are then
selected as the candidate clusters.

\begin{figure}[!ht]
\centering
\includegraphics[height=5cm]{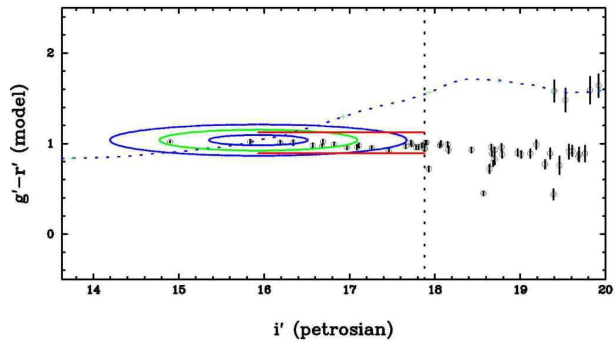}
\caption{SDSS color--magnitude diagram of observed $g - r$ vs. apparent $i$
band for galaxies near a rich cluster at $z$ = 0.15. Ellipses
represent 1, 2, and 3 $\sigma$ contours around the mean BCG color and
magnitude at that redshift. The dotted line indicates the track of BCG
color and magnitude as a function of redshift. The horizontal lines
and vertical dashed line show the region of inclusion for $N_{gal}$
determination. Taken from Hansen \etal (2005).}
\end{figure}

This algorithm appears to be extremely powerful for selecting clusters
in the SDSS.  Simulations suggest that maxBCG recovers and correctly
estimates the richness for greater than 90\% of clusters and groups
present with $N_{gal}\ge 15$ out to $z = 0.3$, with an estimated
redshift dispersion of $\delta z=0.02$. As long as one can obtain a
sufficiently deep photometric catalog, with the appropriate colors to
map the red sequence, this technique can be used to very efficiently
detect clusters. Like all methods that rely on the presence of a red
sequence, it will eventually fail at sufficiently high redshifts,
where the cluster galaxy population becomes more
heterogeneous. However, clusters detected out to $z\sim1-1.5$, even
using non-optical techniques, still show a red sequence, albeit with
larger scatter, which will reduce the efficiency of this
method. Additionally, the definition of $N_{gals}$ as the number of
red sequence galaxies may introduce a bias, as poorer, less
concentrated, or more distant clusters have less well defined
color--magnitude relations, and the luminosity functions for clusters
vary with richness as well (Figure 10 of Hansen \etal 2005).

\subsection{The Cluster Red Sequence Method}

As we have discussed already, the existence of a tight color--magnitude
relation for cluster galaxies provides a mechanism for reducing fore-
and background contamination, enhancing cluster contrast, and
estimating redshifts in cluster surveys. Because the red sequence is
such a strong indicator of a cluster's presence, and is especially
tight for the brighter cluster members, it can be used to detect
clusters to high redshifts ($z\sim1$) with comparatively shallow
imaging, if an optimal set of photometric bands is chosen. This is the
idea behind the Cluster Red Sequence (CRS; Gladders \& Yee 2000)
method, utilized by the Red Sequence Cluster Survey (RCS; Gladders
\etal 2005). Figure 9a shows model color--magnitude tracks for different
galaxy types for $0.1\le z\le1.0$. The cluster ellipticals are the
reddest objects at all redshifts. Even more importantly, if the
filters used straddle the 4000\AA~break at a given redshift, the
cluster ellipticals at that redshift are redder than all galaxies
at all lower redshifts. The only contaminants are more distant, bluer
galaxies, eliminating most of the foreground contamination found in
imaging surveys. The change of the red sequence color with redshift at
a fixed apparent magnitude also makes it a very useful redshift
estimator (L\'{o}pez-Cruz 2004).

\begin{figure}[!ht]
\centering
\leavevmode
\includegraphics[height=6cm]{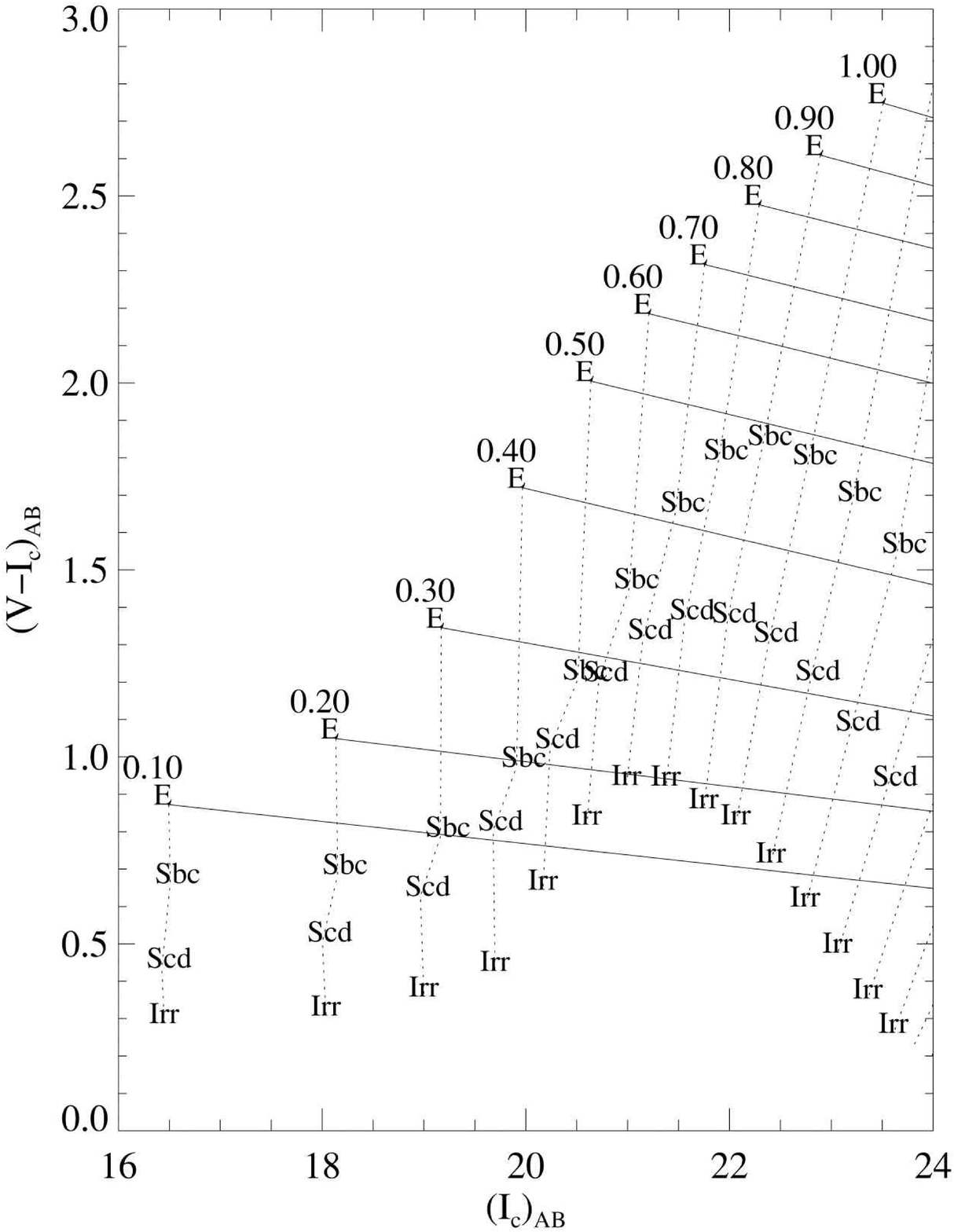}
\includegraphics[height=6cm]{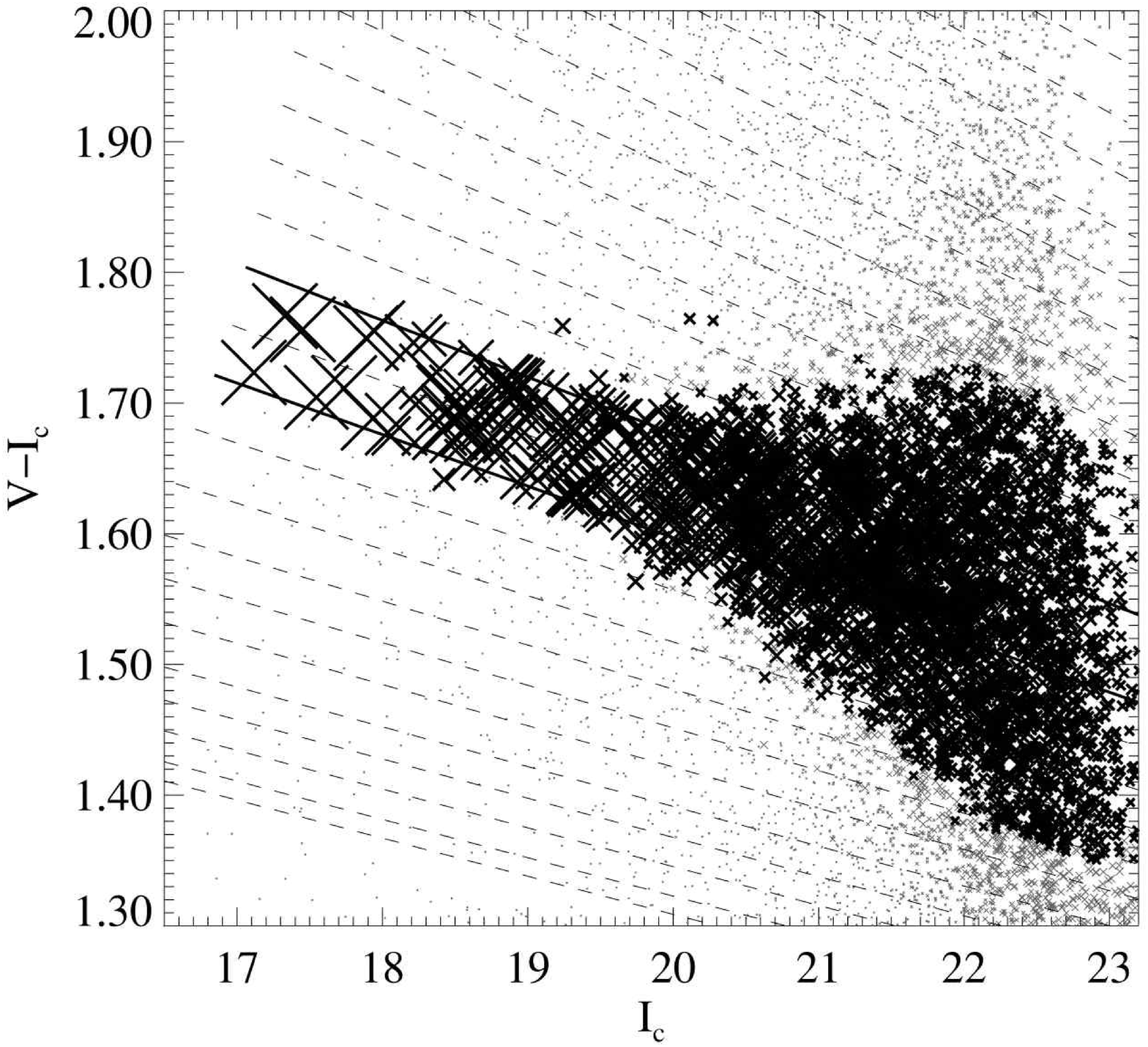}
\caption{{\it Left:} Simulated $(V-I_c)_{AB}$ vs. $(I_c)_{AB}$
color--magnitude diagram. Model apparent magnitudes and colors at
various redshifts for several types of galaxies at a fixed $M_I$ of
-22 are shown. The dotted lines connect galaxies at the same
redshift. Solid near-horizontal lines show the expected slope of the
red sequence at each redshift. {\it Right}: CMD of a CNOC2 Redshift
Survey Patch, with dashed lines showing various color CRS slices. The
galaxy symbols are sized by the probability that they belong to the
color slice defined by the solid lines. Taken from Gladders \& Yee
(2000).}
\end{figure}

Gladders \& Yee generate a set of overlapping color slices based on
models of the red sequence. A subset of galaxies is selected that
belong to each slice, based on their magnitudes, colors, color errors,
and the models. A weight for each chosen galaxy is computed, based on the
galaxy magnitude and the likelihood that the galaxy belongs to the
color slice in question (Figure 9b). A surface density map is then
constructed for each slice using a fixed smoothing kernel, with a
scale radius of 0.33 $h^{-1}$ Mpc. All the slices taken together form
a volume density in position and redshift. Peaks are then selected
from this volume. Gladders \etal 2005 present the results of this
technique applied to the first two RCS patches.

\begin{figure}[!ht]
\centering
\includegraphics[height=6cm]{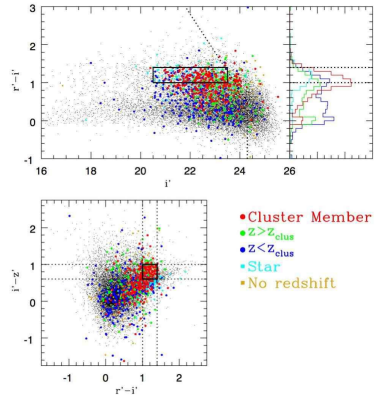}
\caption{$r-i$ vs. $i$ color--magnitude and $r-i$ vs. $i-z$ color--color diagrams for objects in the Cl1604 field.}
\end{figure}

In a similar vein, the High Redshift Large Scale Structure Survey
(Lubin \etal 2006, Gal \etal 2004b, 2005) uses deep multicolor
photometry around known clusters at $z>0.7$ to search for additional
large scale structure. They apply color and color--color cuts to select
galaxies with the colors of spectroscopically confirmed members in the
original clusters. The selected galaxies are used to make adaptive
kernel density maps from which peaks are selected. This technique was
applied to the Cl1604 supercluster at $z\sim0.9$. Starting with two
known clusters with approximately 20 spectroscopic members, there are
now a dozen structures with 360 confirmed members known in this
supercluster. These galaxies typically follow the red sequence, but as
can be seen in Figure 10, the scatter is very large, and many cluster
or supercluster members are actually bluer than the red sequence at
this redshift. Figure 10 shows the $r-i$ vs. $i$ color--magnitude and
$r-i$ vs. $i-z$ color--color diagrams for objects in a $\sim 30'$
square region around the Cl1604 supercluster, with all known cluster
members shown in red. and the color selection boxes marked. Figure 11
shows the density map for this region, with two different significance
thresholds, and the clusters comprising the supercluster marked.
Clearly, in regions such as this, traditional cluster detection
techniques will yield incorrect results, combining multiple clusters,
and measuring incorrect redshifts and richnesses. Figure 12 shows a
3-d map of the spectroscopically confirmed supercluster members,
revealing the complex nature of this structure. Dots are scaled with
galaxy luminosity. While only $\sim 10$ Mpc across on the sky, the
apparent depth of this structure is nearly 10 times greater, making
it comparable to the largest local superclusters.

\begin{figure}[!ht]
\centering
\includegraphics[height=5cm]{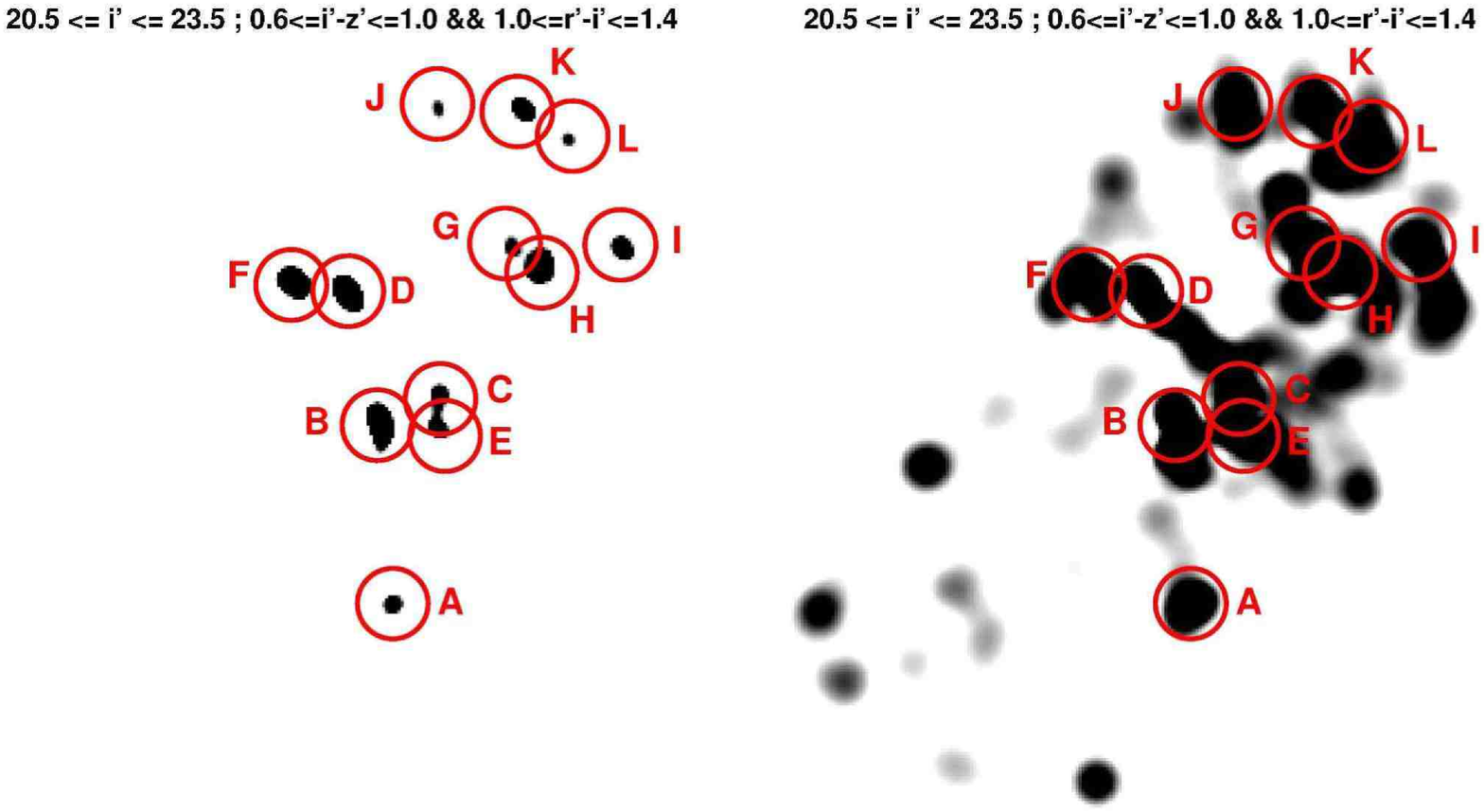}
\caption{Density maps of galaxies meeting the $z\sim0.9$ red galaxy criteria in the Cl1604 field .}
\end{figure}

\begin{figure}[!ht]
\centering
\includegraphics[height=8cm]{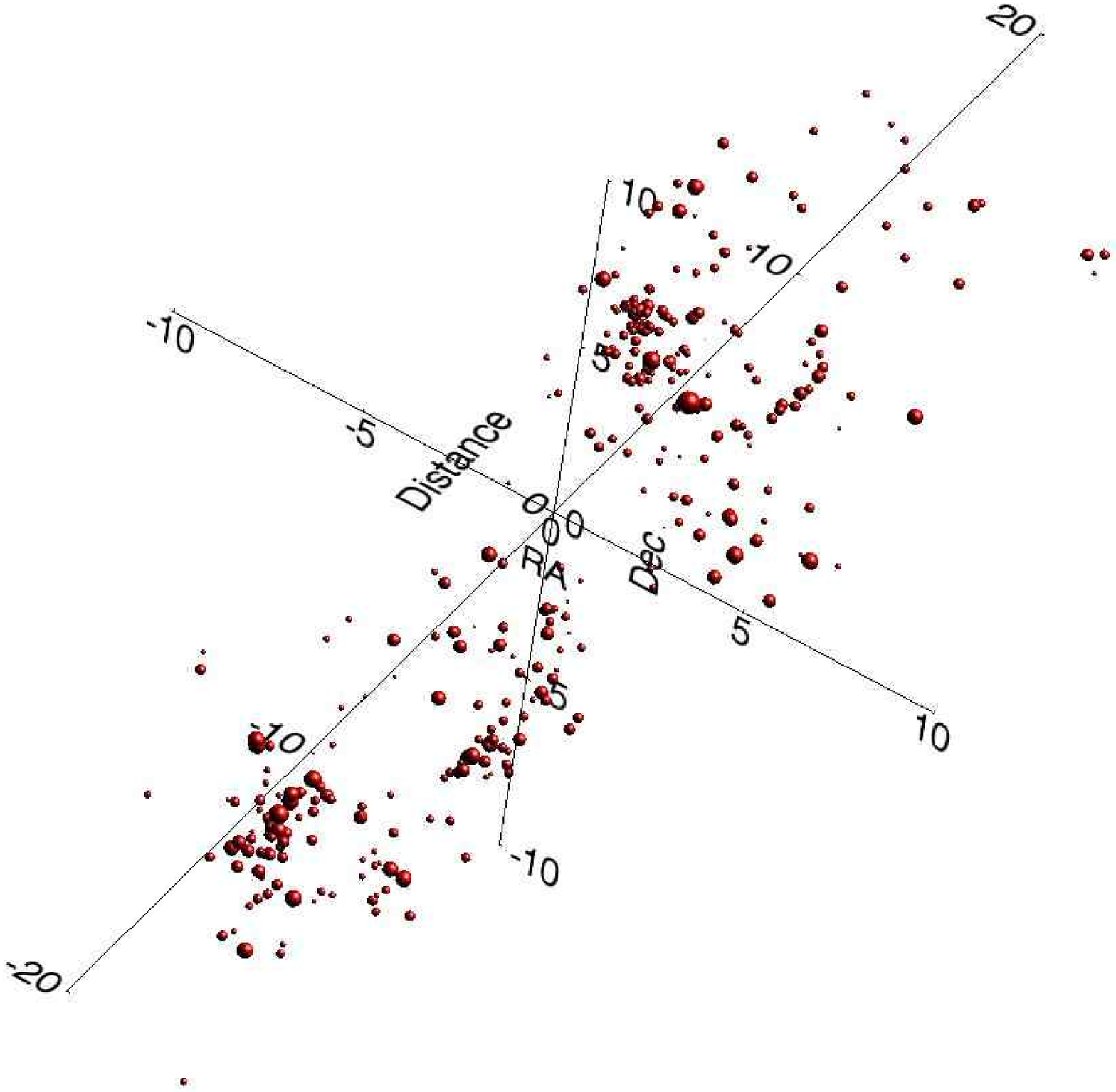}
\caption{Three dimensional spatial distribution of the spectroscopically confirmed Cl1604 supercluster members. Dots are scaled by galaxy luminosity.}
\end{figure}

\section{Conclusions}

It is clear that there exist many methods for detecting clusters in
optical imaging surveys. Some of these are designed to work on very
simple, single--band data (AK, Matched Filter, VT), but will work on
multicolor data as well. Others, such as maxBCG and the CRS method,
rely on galaxy colors and the red sequence to potentially improve
cluster detection and reduce contamination by projections and spurious
objects. Very little work has been done to compare these techniques,
with the exceptions of Kim \etal 2001, Bahcall \etal (2003) and Lopes
\etal (2004), each of whom compared the results of only two or three
algorithms. Even from these tests it is clear that no single technique
is perfect, although some (notably those that use colors) are clearly
more robust. Certainly any program to find clusters in imaging data
must consider the input photometry when deciding which, if any, of
these methods to use.

One of the most vexing issues facing cluster surveys is our inability
to compare directly to large scale cosmological simulations. Most such
simulations are N-body only, but have perfect knowledge of object
masses and positions. Thus, it is possible to construct algorithms to
detect overdensities based purely on mass, but it is {\em not}
possible to obtain the photometric properties of these objects! Recent
work, such as the Millennium Simulation (Springel \etal 2005), is
approaching this goal. It is necessary to extract from these
simulations the magnitudes of galaxies in filters used for actual
surveys, and run the various cluster detection algorithms on these
simulated galaxy catalogs. The results can then be compared to that of
pure mass selection, and the redshift-, structure- and mass-dependent
biases understood. Ideally, this should be done for many large
simulations using different cosmologies, since the galaxy evolution
and selection effects will vary. Such work is fundamental if we are to
use the evolution of the mass function of galaxy clusters for
cosmology. As deeper and larger optical surveys, such as LSST, and
other techniques such as X-ray and Sunyaev--Z'eldovich effect
observations become available, the need for these simulations becomes
ever greater.


\end {document}